\documentclass{amsart}

\usepackage{amsmath,amsfonts,amssymb,amsthm,graphicx,cite}

\newtheorem{theorem}{Theorem}[section]
\newtheorem{lemma}{Lemma}[section]

\newtheorem{cor}{Corollary}[section]
\theoremstyle{remark}
\newtheorem{remark}{Remark}[section]

\newcommand{\ii}{\mathrm{i}}
\newcommand{\ee}{\mathrm{e}}

\newcommand{\Leq}{\preceq}
\newcommand{\Le}{\prec}

\newcommand{\nn}{{a}}
\newcommand{\mm}{{n}}
\newcommand{\kk}{s}

\newcommand{\half}{1/2}

\newcommand{\tH}{{\tilde H}}
\newcommand{\tE}{{\tilde E}}

\newcommand{\HH}{(H)}
\newcommand{\bc}{c}
\newcommand{\pp}[1]{p[{#1}]} 
\newcommand{\ga}{\gamma} 
\renewcommand{\SS}{S} 
\newcommand{\cc}{c}
\newcommand{\PQ}{\check{P}}

\begin{document}

\title[Explicit eigenfunctions of the Calogero model]{Explicit
formulas for the eigenfunctions of the $N$-body Calogero model}

\author{Martin Halln\"as and Edwin Langmann} \address{Mathematical
Physics, KTH Physics, AlbaNova University Center, SE-106 91 Stockholm,
Sweden} \email{martin@theophys.kth.se \textrm{and}
langmann@theophys.kth.se}

\date{\today}

\begin{abstract}
We consider the quantum Calogero model, which describes $N$
non-distinguishable quantum particles on the real line confined by a
harmonic oscillator potential and interacting via two-body
interactions proportional to the inverse square of the inter-particle
distance. We elaborate a novel solution algorithm which allows us to
obtain fully explicit formulas for its eigenfunctions, for arbitrary
coupling parameter and particle number. We also show that our method
applies, with minor changes, to all Calogero models associated with
classical root systems.
\end{abstract}

\maketitle

\section{Introduction}\label{sec1}
In this paper we elaborate a novel solution method for the $N$-body {\em
Calogero model} defined by the Hamiltonian
\begin{equation}\label{hamiltonian}
  H = \sum_{j=1}^N\left(-\partial_{x_j}^2 + x_j^2\right) +
  2\lambda(\lambda-1)\sum_{j<k}\frac{1}{(x_j - x_k)^2},
\end{equation}
where $\lambda>0$ is the coupling parameter, $x_j\in\mathbb{R}$ the
particle coordinates, $\partial_{x_j}:=\partial/\partial x_j$, and
$N=1,2,3,\ldots$ the particle number (we set the harmonic oscillator
frequency $\omega>0$ to 1 without loss of generality: this parameter
can easily be introduced by scaling $x_j\to \sqrt{\omega} x_j$, $H\to
\omega H$, etc.). As is well-known \cite{Cal2,Su2}, this model has
exact eigenfunctions of the form
\begin{equation}\label{factorization}
  \psi_{\mathbf{n}} = \psi_0P_{\mathbf{n}},
\end{equation}
where 
\begin{equation}\label{groundState} 
  \psi_0(\mathbf{x}) =
  \prod_{j=1}^N\ee^{-\frac{1}{2}x_j^2}\prod_{j<k}(x_k -
  x_j)^{\lambda}
\end{equation}
is the groundstate eigenfunction and $P_{\mathbf{n}}(\mathbf{x})$ are
polynomials which are symmetric, i.e., invariant under permutations of
the particle coordinates. These polynomials are labeled by $N$-tuples
$\mathbf{n}=(n_1,\ldots,n_N)$ of non-negative integers,
$n_j\in\mathbb{N}_0$. Due to the permutation symmetry these labels can
be restricted to partitions, i.e.,
\begin{equation*}
  n_1\geq n_2\geq\ldots\geq n_N\geq 0, 
\end{equation*}
but we will {\em not} always make this restriction. The corresponding
exact eigenvalues are given by the following remarkably simple
formulas:
\begin{equation}\label{En}
  E_{\mathbf{n}} = 2(n_1+n_2+\cdots+ n_N) + E_0,\quad E_0 = N(1 +
  \lambda(N - 1)).
\end{equation}
We refer to the $P_{\mathbf{n}}$ as {\em reduced polynomial
eigenfunctions of the Calogero model}, and our aim is to derive
explicit formulas for them. These polynomials are a natural
many-variable generalization of the Hermite polynomials to which they
reduce in the special case $N=1$ \cite{BF}.  Previous results on these
functions \cite{Cal,Su2,Per,Gam,BHV,Kak2,DLM} will be discussed in
more detail below. We also mention that there has been considerable
interest in many-variable generalizations of classical orthogonal
polynomials in the mathematics literature; see e.g.\ \cite{DX,MacD}
and references therein.

Calogero found in his seminal paper \cite{Cal2} the exact eigenvalues
of a closely related model which differs from the one above only in
its center of mass motion (for the convenience of the reader we
discuss the precise relation of these two models in
Appendix~\ref{appA}).  The eigenvalues of the Hamiltonian in
(\ref{hamiltonian}) was given by Sutherland \cite{Su2}, who also
presented an algorithm for constructing the reduced eigenfunctions
$P_{\mathbf{n}}$. This algorithm starts with the ansatz in
(\ref{factorization}), which converts the eigenvalue problem for $H$
into a problem of diagonalizing a certain triangular matrix. Thus, the
eigenvalues of $H$ can be read off from the diagonal of this matrix,
and the eigenfunctions are determined by certain recursion relations
which truncate after a finite number of steps; see Section~\ref{sec2}
for details. To our knowledge, these recursion relations have not been
solved by a closed formula.

In Sutherland's paper \cite{Su2} the emphasis was on a translation
invariant $N$-body model with a $1/\sin^2$-interaction which is also
exactly solvable and the solution algorithm was elaborated in detail
only for this so-called {\em Sutherland model}. In \cite{Lang} one of
us presented an alternative algorithm to solve the Sutherland model
which, different from Sutherland's, also can be generalized to the
elliptic case; see \cite{Lang5} and references therein.  In the
present paper we extend this solution algorithm to the Calogero
model. We stress that the Calogero model is more complicated than the
Sutherland model due to the presence of the harmonic oscillator
potential, and this leads to various interesting and novel features.
It is also interesting to note that, in our approach, the
factorization of the eigenfunctions in (\ref{factorization}) is a
consequence, rather than an essential ingredient, of the
method. Moreover, rather than constructing the eigenfunctions as
linear combinations of the free boson eigenstates as Sutherland, we
obtain a set of somewhat more complicated functions which lead to
simpler recursion relations which we solve explicitly. This gives our
main result: an explicit formula for the reduced polynomial
eigenfunctions of the Calogero model.

We now briefly describe this result.  For each fixed
$\mathbf{x}\in\mathbb{R}^N$ and $\epsilon>0$, let $\mathcal{C}_j$
denote the following set of nested circles in the complex plane:
\begin{equation}\label{Cj}
  \mathcal{C}_j:\; y_j = (\max_{1\leq k\leq N} (|x_k|) + \epsilon
  j)\ee^{\ii\varphi_j},\quad -\pi\leq\varphi_j < \pi,\ j = 1,\ldots,N.
\end{equation}
Using these curves as integration paths, define for
each $\mathbf{n}\in\mathbb{N}_0^N$ the functions
\begin{equation}\label{fn}
  f_{\mathbf{n}}(\mathbf{x}) \, := \prod_{j=1}^N \left(
  \oint_{\mathcal{C}_j} \frac{dy_j}{2\pi\ii y_j} 
  y_j^{n_j}\right) \frac{\prod_{j<k}(1 -
  y_j/y_k)^{\lambda}}{\prod_{j,k=1}^N(1 - x_j/y_k)^{\lambda}}
\end{equation}
which are symmetric polynomials independent of $\epsilon>0$; see
Section~\ref{sec3}.  Our main result is a fully explicit formula for
the functions $P_{\mathbf{n}}$ as linear superpositions of these
functions $f_{\mathbf{n}}$. We use the natural basis elements
$\mathbf{e}_j\in\mathbb{N}_0^N$ defined by $(\mathbf{e}_j)_k : =
\delta_{jk}$ and write $\delta_{\mathbf{n}}(\mathbf{m}):=
\delta_{\mathbf{n},\mathbf{m}}$ for the Kronecker delta.

\begin{theorem}\label{mainthm}
For $\mathbf{n}\in\mathbb{Z}^N$ let
\begin{equation*}
  P_{\mathbf{n}} = \sum_{\mathbf{m}}
  \alpha_{\mathbf{n}}(\mathbf{m})f_{\mathbf{m}}
\end{equation*} 
with the functions $f_{\mathbf{m}}$ defined in (\ref{fn}) and the
coefficients
\begin{equation}\label{alphanm} 
  \begin{split}
    \alpha_{\mathbf{n}}(\mathbf{m}) &= \delta_{\mathbf{n}}(\mathbf{m})
    + \sum_{s=1}^\infty\frac1{4^s s!}\sum_{j_1\leq k_1}\cdots
    \sum_{j_s\leq k_s} \sum_{\nu_1,\ldots,\nu_s=0}^\infty \\&\quad
    \times \delta_{\mathbf{n}}(\mathbf{m} + \mbox{$\sum_{r=1}^{s}
    \mathbf{E}_{j_rk_r}^{\nu_r}$} ) \prod_{r=1}^s g_{j_r
    k_r}(\nu_r;\mathbf{n}
    -\mbox{$\sum_{\ell=1}^r\mathbf{E}_{j_rk_r}^{\nu_r}$}),
  \end{split}
\end{equation}
where we use the shorthand notations
\begin{equation}\label{gjk} 
    g_{jk}(\nu;\mathbf{m}) = \, 2\lambda(\lambda-1) \nu (1-\delta_{jk})
    -\tilde{m}_j(\tilde{m}_j+1)\delta_{\nu 0}\delta_{jk},
\end{equation}
\begin{equation*}
\tilde{m}_j = m_j + \lambda(N+1-j),
\end{equation*}
and
\begin{equation}\label{Ejk}
  \mathbf{E}_{jk}^\nu = (1-\nu)\mathbf{e}_j + (1+\nu) \mathbf{e}_k. 
\end{equation}
Then $P_{\mathbf n}$ is a reduced polynomial eigenfunction of the
Calogero model corresponding to the eigenvalue $E_{\mathbf{n}}$ in
(\ref{En}).
\end{theorem}

\noindent (The proof will be given in Sections~\ref{sec31}--\ref{sec34}.)
 
\medskip

It is important to note that the sums in (\ref{alphanm}) only contain
a finite number non-zero terms. It is also remarkable that
Theorem~\ref{mainthm} is non-trivial already for the simplest case
$N=1$, as discussed in Section~\ref{sec35}. In Section 3.6 we use this
result to construct somewhat more complicated basis functions than the
$f_{\mathbf{n}}$, leading to another explicit formula for the reduced
polynomial eigenfunctions; see Theorem~\ref{altFormTheorem}.

Observe that for $N>1$, this result gives too many eigenfunctions:
they are in Theorem~\ref{mainthm} labeled by elements in
$\mathbb{Z}^N$, but it is known that a complete set of eigenfunctions
can be parameterized by partitions alone. Using the symbolic
programming language MATHEMATICA we have checked for $N=2$ that the
$P_{\mathbf{n}}$ are non-zero eigenfunctions also for non-partitions
$\mathbf{n}$, and we conjecture this to be true for all $N$.  This
over-completeness of our solution poses some interesting questions
discussed in Remark~\ref{rem_overcomplete}.

Similarly as in Sutherland's algorithm \cite{Su2}, we obtain the
coefficients $\alpha_{\mathbf{n}}(\mathbf{m})$ by diagonalizing a
certain triangular matrix, and they are therefore non-zero only for
$\mathbf{m}\Leq\mathbf{n}$ in some partial ordering $\Leq$; see
Section~\ref{sec2}. However, this partial ordering is different from
Sutherland's, and the matrix we get is simpler, which is why we can
find its explicit eigenvectors.

As found by Olshanetsky and Perelomov \cite{OP}, the model discussed
so-far can be naturally associated with the root system $A_{N-1}$, and
there are exactly solvable variants of the Calogero model related to
all other root systems; see \cite{OP2} for a comprehensive review. In
particular, the Calogero models associated with the remaining
classical root systems \cite{OP2} can all be brought to the form of
the $B_N$ Hamiltonian
\begin{equation}\label{HBN} 
  \begin{split}
    H_{B_N} = & \sum_{j=1}^N\left( -\partial_{x_j}^2 + x_j^2 +
    \frac{\mu(\mu - 1)}{x_j^2}\right) + 4\lambda(\lambda - 1)\sum_{
    j<k}\frac{(x_j^2+x_k^2)}{(x_j^2 - x_k^2)^2}
  \end{split}
\end{equation}
with two coupling parameters $\mu,\lambda>0$. To demonstrate the
generality of our solution method we show that the construction of
eigenfunctions with minor changes goes through also in this case, and
we thereby obtain explicit formulas for a many-variable generalization
of the Laguerre polynomials. This adds support to our hope that the
method can be used so solve any Calogero-Sutherland type model. We
should mention that the $B_N$ Calogero model also can be solved using
Sutherland's method, of course.

As mentioned in the first paragraph, various other explicit results
for the reduced polynomial eigenfunctions of the Calogero model exist
in the literature. Calogero obtained such results for the cases $N =
2,3$ \cite{Cal}. By exploiting an underlying group structure of the
Hamiltonian, Perelemov \cite{Per} for $N=4$ and Gambardella for $N=5$
\cite{Gam} obtained the eigenfunctions in terms of ``raising''
operators acting on the groundstate. More recently these operator
solutions were generalized to all $N$ \cite{BHV,Kak2}. We also mention
that Desrosiers et.al.\ obtained explicit results for the
eigenfunctions of a supersymmetric generalization of the Calogero
model using a determinantal construction \cite{DLM}. Our results seem
different and complementary to these.

The plan of the rest of this paper is as follows. In
Section~\ref{sec2} we fix our notation and shortly review
Sutherland's solution of the Calogero model \cite{Su2} and a simple
variant thereof which, as we argue, is somewhat more natural. In
Section~\ref{sec3} we present our solution of the $A_{N-1}$ Calogero
model and thereby prove Theorem~\ref{mainthm}. We also comment on the
one-particle case, and we sketch a variant of our solution method
which provides another explicit formula for the eigenfunctions. Our
solution of the $B_N$ Calogero model is presented in
Section~\ref{sec4}. We end with a few concluding remarks in
Section~\ref{sec5}. Some technical details are deferred to two
appendices.

\section{Sutherland's solution algorithm}\label{sec2}
In this section we fix our notation and, to put our work into context,
briefly review Sutherland's solution of the Calogero model
\cite{Su2}. We will actually discuss a somewhat simpler variant of
this solution method, as explained below.

In the discussion below we make use of some notational conventions
from the theory of partitions which we now recall; see e.g.\
\cite{MacD}. For partitions $\mathbf{n} = (n_1,n_2,\ldots,n_N)$, the
non-zero $n_i$ are called the {\em parts} of $\mathbf{n}$, and we use
the short hand notation
\begin{equation*}
|\mathbf{n}| \, : = n_1 + n_2 + \ldots + n_N.
\end{equation*}
We also introduce a partial ordering of partitions: for two partitions
$\mathbf{m},\mathbf{n}$ we write
\begin{equation*}
  \mathbf{m}\leq\mathbf{n}\; \Leftrightarrow \; \sum_{k=1}^j m_k \leq
  \sum_{k=1}^j n_k \quad \forall j = 1,\ldots,N. \label{leq}
\end{equation*} 
We will furthermore write $\mathbf{m}<\mathbf{n}$ if
$\mathbf{m}\leq\mathbf{n}$ and $|\mathbf{m}|\neq |\mathbf{n}|$.

The starting point of Sutherland's algorithm is the observation that
the function $\psi_0$ in (\ref{groundState}) is the groundstate of the
Hamiltonian $H$ in (\ref{hamiltonian}), a fact which can be proved by
a straightforward computation; see Remark~\ref{rem2}. As previously
mentioned, another key insight is that any eigenfunction of the
Hamiltonian $H$ in (\ref{hamiltonian}) can be factorized into a
symmetric polynomial and the groundstate.  This implies that such a
symmetric polynomial is an eigenfunction of the differential operator
\begin{equation}\label{reducedHam}
  \tH :=\ \psi_0^{-1}H\psi_0 - E_0 = \sum_{j=1}^N\left(
  -\partial_{x_j}^2 + 2 x_j\partial_{x_j}\right) -
  2\lambda\sum_{j<k}\frac{1}{x_j - x_k}(\partial_{x_j} -
  \partial_{x_k}).
\end{equation}
The idea is now to construct these polynomials as linear combinations
of the monomials
\begin{equation}
  M_{\mathbf{n}} = \sum_{P\in S_N}x_{P(1)}^{n_1}\cdots x_{P(N)}^{n_N},
\label{Mn}
\end{equation}
where $\mathbf{n}$ is a partition of length $N$ and $S_N$ the
permutation group of $N$ elements. We note in passing that the
standard normalization of these monomials is different in that the sum
in (\ref{Mn}) is restricted to the distinct permutations of the parts
$n_j$ (see e.g.\ \cite{MacD}), but for our purposes the normalization
where one sums over all permutations is more convenient. To proceed we
use the fact that
\begin{equation*}
  (-\partial_x^2 + 2x\partial_x) x^n = 2n x^n - n(n-1)x^{n-2} 
\end{equation*}
as well as the identity
\begin{equation*}
\begin{split} 
  \frac{1}{x-y}(\partial_x - \partial_y)(x^ny^m + y^nx^m) = (n -
  m)\sum_{k=1}^{n-m-1}x^{n-1-k}y^{m-1+k}\\ - m ( x^{n-1}y^{m-1} +
  y^{n-1}x^{m-1} ),
\end{split}
\end{equation*}
valid for all $x,y\in\mathbb{R}$ and $m,n\in\mathbb{N}_0$ such that
$n\geq m$. A proof of this identity can be found in Appendix B. It
follows that
\begin{equation}\label{action}
  \begin{split}
    \tH M_{\mathbf{n}} = 2|\mathbf{n}|M_{\mathbf{n}} -
    \sum_{j=1}^N n_j(n_j-1) M_{\mathbf{n}-2\mathbf{e}_j} \\ -
    \lambda\sum_{j<k} \sum_{\nu=0}^{\lfloor \frac{n_j-n_k}2\rfloor}
    (2-\delta_{2\nu,n_j-n_k})\left( (1-\delta_{\nu,0}) n_j -
    n_k\right) M_{\mathbf{n} - (\nu+1)\mathbf{e}_j+(\nu-1)
    \mathbf{e}_k}
  \end{split} 
\end{equation}
where $\lfloor n/2 \rfloor=n/2$ or $(n-1)/2$ for even or odd integers
$n$, respectively. Hence, the action of $\tH$ on the monomials
$M_{\mathbf{n}}$ has triangular structure in the following sense:
\begin{equation*}
  \tH M_{\mathbf{n}} = 2|\mathbf{n}|M_{\mathbf{n}} +
  \sum_{\mathbf{m} < \mathbf{n}}
  b_{\mathbf{n}\mathbf{m}}M_{\mathbf{m}}\label{Hpm1}
\end{equation*}
for certain coefficients $b_{\mathbf{n}\mathbf{m}}$ which can be
determined from (\ref{action}). This suggests that $\tH$ has
eigenfunctions of the form
\begin{equation}\label{Pn}
  P_{\mathbf{n}} = M_{\mathbf{n}} + \sum_{\mathbf{m}<\mathbf{n}}
  u_{\mathbf{n}\mathbf{m}} M_{\mathbf{m}}
\end{equation}
with corresponding eigenvalues $\tilde
E_{\mathbf{n}}=2|\mathbf{n}|$. Indeed, inserting this result into the
Schr\"odinger equation $\tH P_{\mathbf{n}}= \tilde
E_{\mathbf{n}}P_{\mathbf{n}}$ and using the fact that the monomials
$M_{\mathbf{m}}$ are linearly independent we obtain the following
system of equations:
\begin{equation}\label{rec}
  (\tE_{\mathbf{n}}-\tE_{\mathbf{m}})u_{\mathbf{n}\mathbf{m}} =
  b_{\mathbf{n}\mathbf{m}} + \sum_{\mathbf{m}<\mathbf{k}<\mathbf{n}}
  u_{\mathbf{n}\mathbf{k}}b_{\mathbf{k}\mathbf{m}},\quad
  \mathbf{m}<\mathbf{n}.
\end{equation}
It is important to note that $|\mathbf{m}|<|\mathbf{n}|$ for all
$\mathbf{m}<\mathbf{n}$.  Moreover, for each partition $\mathbf{n}$,
there exists only a finite number of partitions
$\mathbf{m}<\mathbf{n}$. Thus, (\ref{rec}) gives a well-defined
recursion procedure for computing all coefficients
$u_{\mathbf{n}\mathbf{m}}$ in a finite number of steps.

As mentioned, the method described above is a somewhat simpler variant
of Sutherland's original method \cite{Su2} who, instead of the
monomials $M_{\mathbf{n}}$, used somewhat more complicated basis
functions which we now describe. Let
\begin{equation}
  H_n(x) = \sum_{k=0}^{\lfloor
  n/2\rfloor}(-1)^k\frac{n!}{k!(n-2k)!}(2x)^{n-2k}\label{Hermite}
\end{equation}
denote the Hermite polynomial of order $n\in\mathbb{N}_0$, satisfying
the differential equation
\begin{equation*}
  (-\partial_x^2 + 2 x\partial_x)H_n(x)=2n  H_n(x). 
\end{equation*}
Let $M^{\HH}_{\mathbf{n}}$ denote the symmetric polynomial
\begin{equation*}
  M^{\HH}_{\mathbf{n}}(\mathbf{x}) =
  M_{\mathbf{n}}(H_{n_1}(x_1),\ldots,H_{n_N}(x_N)) = \sum_{P\in S_N}
  H_{n_1}(x_{P(1)})\ldots H_{n_N}(x_{P(N)}).
\end{equation*}
These symmetric polynomials are obviously eigenstates of the
differential operator $\tH$ in (\ref{reducedHam}) for the free
case $\lambda=0$ with eigenvalues $2|\mathbf{n}|$. The key identity is
\cite{Su2}
\begin{equation}\label{hermId}
  \begin{split}
    \frac{1}{x-y}(\partial_x - \partial_y)\left(  H_n(
    x)H_m(y) + H_n(y)H_m(
    x)\right)\\ = \sum_{r=1}^n\sum_{s=1}^m \bc_{rs}(n,m)\left( 
    H_{n-r}(x)H_{m-s}(y) + H_{n-r}(
    y)H_{m-s}(x)\right)
  \end{split}
\end{equation}
for all $n,m\in\mathbb{N}_0$ and certain real coefficients
$\bc_{rs}$. Since a proof of this identity is not contained in
Sutherland's paper \cite{Su2} we provide a sketch thereof in
Appendix~\ref{appB}. This identity shows that the action of $\tH$ on
the symmetric polynomials $M^{\HH}_{\mathbf{n}}$ is triangular, which
suggests that there are eigenfunctions $P_{\mathbf{n}} =
M^{\HH}_{\mathbf{n}} + \sum_{\mathbf{m}<\mathbf{n}}
v_{\mathbf{n}\mathbf{m}}M^{\HH}_{\mathbf{m}}$ of $\tH$ with
eigenvalues $2|\mathbf{n}|$ and a recursive procedure to compute all
coefficients $v_{\mathbf{n}\mathbf{m}}$ from
$v_{\mathbf{n}\mathbf{n}}=1$, as above. However, the explicit formulas
for the coefficients $\bc_{rs}(n,m)$ were not provided in \cite{Su2},
and they indeed seem rather difficult to obtain: we neither found them
in the literature, nor where we able to derive them.

It is interesting to note that the recursion relations in (\ref{rec})
can be inverted to yield explicit formulas for the coefficients
$u_{\mathbf{n}\mathbf{m}}$: introducing $u_{\mathbf{n}\mathbf{n}}= 1$
and a linear operator $R$ acting on these coefficients as follows:
\begin{equation*}
  Ru_{\mathbf{n}\mathbf{m}} = \frac{1-\delta_{\mathbf{n},\mathbf{m}}}{
  \tE_{\mathbf{n}}-\tE_{\mathbf{m}}} \left( b_{\mathbf{n}\mathbf{m}} +
  \sum_{\mathbf{m}<\mathbf{k}<\mathbf{n}}u_{\mathbf{n}\mathbf{k}}
  b_{\mathbf{k}\mathbf{m}}\right) ,
\end{equation*}
we can rewrite the recursion relations as
\begin{equation*}
  u_{\mathbf{n}\mathbf{m}} = \delta_{\mathbf{n},\mathbf{m}} +
  Ru_{\mathbf{n}\mathbf{m}}.
\end{equation*}
It follows that they can be inverted according to
\begin{equation*}
  u_{\mathbf{n}\mathbf{m}} = (1-R)^{-1}\delta_{\mathbf{n},\mathbf{m}} =
  \sum_{s=0}^{\infty}R^s\delta_{\mathbf{n},\mathbf{m}},
\end{equation*}
where it is important to note that the expansion of the geometric
series is well-defined since it only contains a finite number of
non-zero terms; see below. From the definition of the linear operator
$R$ given above and the fact that $\tE_{\mathbf{n}}- \tilde
E_{\mathbf{m}}=2(|\mathbf{n}|-|\mathbf{m}|)$ now follows that
\begin{equation}\label{explicit} 
  u_{\mathbf{n}\mathbf{m}} = \delta_{\mathbf{n},\mathbf{m}} +
  \frac{1-\delta_{\mathbf{n},\mathbf{m}}}{2(|\mathbf{n}-\mathbf{m}|)}
  \left(b_{\mathbf{n}\mathbf{m}} + \sum_{s=1}^{\infty}
  \sum_{\mathbf{m}<\mathbf{k}_1<\cdots<\mathbf{k}_s<\mathbf{n}}
  \frac{b_{\mathbf{n}\mathbf{k}_1}b_{\mathbf{k}_1\mathbf{k}_2} \ldots
  b_{\mathbf{k}_s\mathbf{m}} }{2^s
  \prod_{r=1}^{s}(|\mathbf{n}|-|\mathbf{k}_r|)}\right).
\end{equation}
The restrictions imposed by the inequality in the second sum clearly
implies that this series representation for the coefficients
$u_{\mathbf{n}\mathbf{m}}$ only contains a finite number of non-zero
terms. Also note that each term is well-defined. However, this formula
is not very useful since the $b_{\mathbf{n}\mathbf{m}}$ are not given
by a simple formula. Indeed, to deduce these latter coefficients from
(\ref{action}) it is important to note that $\mathbf{n}-2\mathbf{e}_j$
is {\em not}, in general, a partition, e.g.\
$(3,2,2)-2\mathbf{e}_1=(1,2,2)$. A similar remark applies to the last
term in (\ref{action}). We therefore implicitly used an extension of
the monomials $M_{\mathbf{n}}$ to non-partitions. To make this precise
we introduce an ordering symbol as follows: for each
$\mathbf{a}\in\mathbb{N}_0^N$ we let $\pp{\mathbf{a}}$ denote the
corresponding partition obtained by permuting the elements of
$\mathbf{a}$, e.g., $\pp{(3,1,0,4)}=(4,3,1,0)$. We can then define
$M_{\mathbf{a}}:= M_{\pp{\mathbf{a}}}$, which naturally extends the
definition of the monomials $M_{\mathbf{n}}$ to non-partitions.  Using
this definition we deduce from (\ref{action}) that
\begin{equation*}
  \begin{split}
    b_{\mathbf{n}\mathbf{m}} &= -\sum_{j=1}^N n_j(n_j-1)
    \delta_{\pp{\mathbf{n}-2\mathbf{e}_j},\mathbf{m} } -
    \lambda\sum_{j<k} \sum_{\nu=0}^{\lfloor
    \frac{n_j-n_k}{2}\rfloor} \\&\quad \times
    (2-\delta_{2\nu,n_j-n_k})\left( (1-\delta_{\nu,0})n_j - n_k\right)
    \delta_{\pp{\mathbf{n} - (\nu+1)\mathbf{e}_j+(\nu-1)
    \mathbf{e}_k},\mathbf{m}} .
  \end{split} 
\end{equation*}
Inserting this in (\ref{explicit}) one hopes that, due to the
Kronecker deltas, the sums simplify considerably. However, the
appearance of the ordering symbol $\pp{\cdot}$ makes the resulting
formula awkward to use.  We therefore conclude that the Sutherland
algorithm does not lead to simple explicit formulas for the
eigenfunctions.  The same difficulty arises in Sutherland's original
algorithm described above.

\section{Alternative solution algorithm}\label{sec3}
In this section we present our alternative method for solving the
Calogero model defined by the Hamiltonian in (\ref{hamiltonian}) and,
in particular, prove Theorem \ref{mainthm}.

We will to a large extent use the notation introduced in the beginning
of Section~\ref{sec2}, with the important difference that elements
$\mathbf{m},\mathbf{n}\in\mathbb{Z}^N$ now will be ordered as follows:
\begin{equation*}
  \mathbf{m}\Leq\mathbf{n}\; \Leftrightarrow \; \sum_{k=N+1-j}^N m_k
  \leq \sum_{k=N+1-j}^N n_k, \quad \forall j = 1,\ldots,N.
\end{equation*}

\subsection{A remarkable identity}\label{sec31}
We start by proving a particular functional identity, which is the
starting point for our construction.

\begin{lemma}\label{FIdLemma}
Let $c_N=2(1-\lambda)N$ and
\begin{equation*}
  F(\mathbf{x},\mathbf{y}) = \frac{\prod_{j=1}^N
  \ee^{-\frac{1}{2}(x_j^2-y_j^2)}\prod_{j<k}(x_k -
  x_j)^{\lambda}(y_k - y_j)^{\lambda}}{\prod_{j,k=1}^N(y_k -
  x_j)^{\lambda}}.
\end{equation*}
Then
\begin{equation}
\label{FId}
  H(\mathbf{x})F(\mathbf{x},\mathbf{y}) =
  \lbrack H(\mathbf{y}) + c_N\rbrack F(\mathbf{x},\mathbf{y}),
\end{equation}
where $H=H(\mathbf{x})$ is the Hamiltonian in (\ref{hamiltonian}) and
similarly for $H(\mathbf{y})$.
\end{lemma}
\begin{proof} 
We set $\mathcal{N}=2N$, $X_j=x_j$, $X_{N+j}=y_j$, $m_j=+1$ and 
$m_{N+j}=-1$ for $j=1,2,\ldots,N$. Then $H(\mathbf{x})-
H(\mathbf{y})=\mathcal{H}(\mathbf{X})$ with
\begin{equation}
\label{calH} 
  \mathcal{H} = \sum_{j=1}^{\mathcal{N}}\left(- \frac{1}{m_j}
  \partial_{X_j}^2 + m_j X_j^2\right) + \sum_{j<k}\frac{\lambda
  (m_j+m_k)(\lambda m_jm_k-1)}{(X_j-X_k)^2}
\end{equation}
and $F(\mathbf{x},\mathbf{y})=\Psi_0(\mathbf{X})$ with
\begin{equation}
\label{Psi0} 
  \Psi_0(\mathbf{X}) = \prod_{j=1}^\mathcal{N} \ee^{-\frac12 m_j
  X_j^2} \prod_{j<k } (X_k-X_j)^{m_jm_k\lambda}.
\end{equation}
To prove the lemma will we show by explicit computation that
\begin{equation}
\label{ID}
  (\mathcal{H}- \mathcal{E}_0) \Psi_0(\mathbf{X}) = 0 
\end{equation}
with the constant 
\begin{equation}
  \mathcal{E}_0= \lambda\left(\sum_{j=1}^{\mathcal{N}} m_j\right)^2 +
  \sum_{j=1}^{\mathcal{N}} (1-\lambda m_j^2) \label{calE0}
\end{equation}
and note that $\mathcal{E}_0=c_N$.  For that we introduce the operator
\begin{equation*}
  \mathcal{D} = \sum_{j=1}^{\mathcal{N}} \frac1{m_j} Q_j^+ Q_j^-
\end{equation*}
with
\begin{equation*}
  Q^\pm_j = \pm\partial_{X_j} + W_j ,\quad W_j = -m_j X_j + \sum_{k\neq
  j} \frac{\lambda m_jm_k}{(X_j-X_k)} .
\end{equation*}
Note that $Q^-_j\Psi_0=0$ for all $j=1,\ldots,\mathcal{N}$, and hence
that
\begin{equation*}
\mathcal{D}\Psi_0 = 0. 
\end{equation*}
This implies the identity in (\ref{ID}) since $\mathcal{D} =
\mathcal{H}-\mathcal{E}_0$, as can be shown by straightforward
computations. Indeed,
\begin{equation*}
  \mathcal{D} = \sum_{j=1}^N \frac1{m_j}\left( - \partial_{X_j}^2 + W_j^2 +
  (\partial_{X_j} W_j) \right) = \mathcal{H} - \mathcal{R},
\end{equation*}
where the reminder terms 
\begin{equation*}
  \mathcal{R} = \sum_{j=1}^N 1 + 2\sum_{k\neq j} \lambda m_j m_k
  \frac{X_j}{X_j-X_k} + \sum_{\substack{k,\ell\neq j\\ \ell\neq k}}
  \frac{\lambda m_j m_k m_\ell}{(X_k-X_j)(X_j-X_\ell)}
\end{equation*}
add up to the constant $\mathcal{E}_0$: upon symmetrization the double
sum becomes independent of the $X_j$ and equal to $\sum_{j\neq k}
\lambda m_j m_k = \lambda (\sum_j m_j )^2-\lambda \sum_j m_j^2$ and
the triple sum vanishes, as can be seen by symmetrizing in the
indices $j,k,\ell$ and using the identity
\begin{equation*}
  \frac1{(X_k-X_j)(X_j-X_\ell)} + \frac1{(X_k-X_j)(X_\ell-X_k)} +
  \frac1{(X_j-X_\ell)(X_\ell-X_k)} = 0.
\end{equation*}
\end{proof}

\begin{remark}\label{rem2} 
It is easy to see that we have, in fact, proved a more general result:
the identity in (\ref{ID}) holds true for all $\mathcal{N}=2,3,\ldots$
and arbitrary real parameters $m_j$. Obviously, one particular
consequence of this latter result is the fact that the function
$\psi_0$ in (\ref{groundState}) is the groundstate of the Hamiltonian
$H$ in (\ref{hamiltonian}). It is not difficult to see that we have
proved a similar fact for a more general case: {\em if all $m_j>0$
then $\mathcal{H}$ in (\ref{calH}) defines a self-adjoint operator on
the Hilbert space $L^2(\mathbb{R}^{\mathcal{N}})$ with
$\Psi_0(\mathbf{X})$ in (\ref{Psi0}) as groundstate}. This is true
since $Q_j^+$ then is the Hilbert space adjoint of $Q_j^-$, implying
that $\mathcal{D}=\mathcal{H}-\mathcal{E}_0$ is a sum of non-negative
terms $(Q_j^-)^*Q_j^-/m_j$ and thus defines a unique non-negative
self-adjoint operator via the Friedrichs extension; see e.g.\
\cite{RS2}.  We thus recover a known generalization of the Calogero
model where the particles can have different masses $m_j>0$ and such
that its exact groundstate and groundstate energy can be computed
exactly \cite{Forr,MMS}. Other interesting special cases will be
discussed in Remark~\ref{remspc}.
\end{remark}

\subsection{Integral transformation}\label{sec32}
The idea is now to apply to the identity in (\ref{FId}) an integral
transform $ \prod_{j=1}^N \left( \oint_{\mathcal{C}_j} dy_j (2\pi\ii
y_j)^{-1}\phi_{n_j}(y_j) \right) $ with the integration paths in
(\ref{Cj}) and certain functions $\phi_{j}(y_j)$ to be chosen such
that this transform is well-defined. We observe that
\begin{equation*}
  F(\mathbf{x},\mathbf{y}) = \psi_0(\mathbf{x}) \prod_{j=1}^N\left(
  \ee^{\frac{1}{2} y_j^2} y_j^{\lambda(j-N-1) }\right)
  \frac{\prod_{j<k} (1 - y_j/y_k)^{\lambda}}{\prod_{j,k=1}^N(1 -
  x_j/y_k)^{\lambda}},
\end{equation*}
which shows that if we choose
\begin{equation} 
  \phi_{j}(y_j) = \ee^{-\frac12 y_j^2} y_j^{\tilde n_j} , \quad \tilde
  n_j =n_j + \lambda(N+1-j) \label{ntilde}
\end{equation} 
with integers $n_j$, then this transformation is well-defined for all
$\lambda>0$. Indeed, for all ${\mathbf{n}}\in\mathbb{Z}^N$,
\begin{equation*} 
  \begin{split}
    \hat F_{\mathbf{n}}(\mathbf{x}) := \prod_{j=1}^N \left(
    \oint_{\mathcal{C}_j} \frac{dy_j}{2\pi\ii y_j} \ee^{-\frac12
    y_j^2} y_j^{\tilde n_j} \right) F(\mathbf{x},\mathbf{y}) =
    \psi_0(\mathbf{x}) f_{\mathbf{n}}(\mathbf{x})
  \end{split} 
\end{equation*}
with $f_{\mathbf{n}}$ the functions defined in (\ref{fn}).

The application of this integral transform to the l.h.s.\ of the
identity in (\ref{FId}) obviously gives $H\hat
F_{\mathbf{n}}(\mathbf{x})$. To compute the integral transform of the
r.h.s.\ we observe that
\begin{equation*}
  (-\partial_{y_j}^2 + y_j^2) \ee^{-\frac12 y_j^2} y_j^{\tilde n_j-1}
  = \ee^{-\frac12 y_j^2} \left( (2\tilde n_j-1) y_j^{\tilde n_j-1}
  -(\tilde n_j-1)(\tilde n_j-2) y_j^{\tilde n_j-3}\right)
\end{equation*}
and that
\begin{equation*}
  \frac1{(y_j- y_k)^2} = \frac1{y_k^2(1- y_j/y_k)^2} =
  \sum_{\nu=1}^\infty \nu y_j^{\nu-1} y_k^{-\nu-1}
\end{equation*} 
for all $|y_j|<|y_k|$. Using these two facts and the shorthand
notation
\begin{equation*}
  \ga = 2\lambda(\lambda-1)
\end{equation*}
we obtain by straightforward computations that
\begin{equation}\label{HFn}
  H\hat F_{\mathbf{n}} = E_{\mathbf{n}}\hat F_{\mathbf{n}}
  -\sum_{j=1}^N (\tilde n_j-1)(\tilde n_j-2) \hat F_{\mathbf{n}-2
  \mathbf{e}_j} + \ga \sum_{j<k} \sum_{\nu=1}^\infty\nu \hat
  F_{\mathbf{n} - (1-\nu) \mathbf{e}_j - (1+\nu) \mathbf{e}_k},
\end{equation} 
where we used that $\sum_{j=1}^N 2(2\tilde n_j-1) + c_N =
E_{\mathbf{n}}$, as given in (\ref{En}).

\subsection{Construction of eigenfunctions}\label{sec33}
Equation (\ref{HFn}) shows that the action of $H$ on the functions
$\hat F_{\mathbf{n}}$ has triangular structure: $H\hat F_{\mathbf{n}}$
is a linear combination of functions $\hat F_{\mathbf{m}}$ with
$\mathbf{m}\Leq \mathbf{n}$. Similarly as in the Sutherland algorithm,
this suggests that the Calogero model has eigenfunctions of the form
\begin{equation}\label{psin}
  \psi_{\mathbf{n}} = \alpha_{\mathbf{n}}(\mathbf{n})\hat
  F_{\mathbf{n}} + \sum_{\mathbf{m}\prec\mathbf{n}}
  \alpha_{\mathbf{n}}(\mathbf{m}) \hat F_{\mathbf{m}}
\end{equation}
with eigenvalues $E_{\mathbf{n}}$ and certain coefficients
$\alpha_{\mathbf{n}}(\mathbf{m})$. Indeed, inserting this formula for
$\psi_{\mathbf{n}}$ in (\ref{HFn}), we obtain by straightforward
computations that 
\begin{equation*}
  \begin{split}
    H\psi_{\mathbf{n}} = \sum_{\mathbf{m}\preceq\mathbf{n}} \Biggl(
    E_{\mathbf{m}} \alpha_{\mathbf{n}}(\mathbf{m}) - \sum_{j=1}^N
    (\tilde m_j+1)\tilde{m}_j\alpha_{\mathbf{n}}(\mathbf{m}+2\mathbf{e}_j) +
    \\ \ga \sum_{j<k} \sum_{\nu=1}^\infty \nu
    \alpha_{\mathbf{n}}(\mathbf{m} + (1-\nu) \mathbf{e}_j + (1+\nu)
    \mathbf{e}_k) \Biggr) \hat F_{\mathbf{m}} . 
  \end{split}
\end{equation*}
We conclude that the validity of the Schr\"odinger equation
$H\psi_{\mathbf{n}} = E_{\mathbf n}\psi_{\mathbf{n}}$ is implied by
the recursion relations
\begin{equation}
\label{EEal}
  2(|\mathbf{n}| - |\mathbf{m}|) \, \alpha_{\mathbf{n}}(\mathbf{m}) =
  \sum_{j\leq k} \sum_{\nu=0}^\infty
  g_{jk}(\nu;\mathbf{m}) \alpha_{\mathbf{n}}(\mathbf{m} +
  \mathbf{E}_{jk}^{\nu}),
\end{equation}
with $g_{jk}(\nu;\mathbf{m})$ and $E_{jk}^\nu$ defined in (\ref{gjk})
and (\ref{Ejk}), respectively; we used $E_{\mathbf n} - E_{\mathbf m}
= 2(|{\mathbf n}|-|{\mathbf m}|)$.  We now construct an explicit
solution of (\ref{EEal}). The triangular structure of the
eigenfunctions implies that $\alpha_{\mathbf{n}}(\mathbf{m})=0$ unless
$\mathbf{m}\prec\mathbf{n}$ or $\mathbf{m}=\mathbf{n}$, and that we
can set
\begin{equation*}
  \alpha_{\mathbf{n}}(\mathbf{n})=1
\end{equation*}
without loss of generality. This implies that the recursion relations
in (\ref{EEal}) can be written as follows:
\begin{equation*}
  \alpha_{\mathbf{n}} = \delta_{\mathbf{n}} + \SS \alpha_{\mathbf{n}},
\end{equation*}
where the operator $\SS$ is defined by 
\begin{equation}\label{Rn}
  (\SS\alpha_{\mathbf{n}})(\mathbf{m}) :=
  \frac{1-\delta_{\mathbf{n}}(\mathbf{m})}{2(|{\mathbf n}|-|{\mathbf
  m}|)} \sum_{j\leq k} \sum_{\nu=0}^\infty
  g_{jk}(\nu;\mathbf{m}) \alpha_{\mathbf{n}}(\mathbf{m} +
  \mathbf{E}_{jk}^\nu )
\end{equation}
for $\mathbf{m}\Le\mathbf{n}$, which allows us to suppress the common
argument $\mathbf{m}$.  This later equation can now be solved to yield
\begin{equation*}
  \alpha_{\mathbf{n}} = (1-\SS)^{-1} \delta_{\mathbf{n}} =
  \sum_{s=0}^\infty \SS^s \delta^{\phantom s}_{\mathbf{n}},
\end{equation*}
where the latter expansion of the geometric series is well-defined
since it only contains a finite number of non-zero terms, as shown
below. Using (\ref{Rn}) we deduce that
\begin{equation*}
  \begin{split}
    (\SS^s_{\mathbf{n}} \delta^{\phantom s}_{\mathbf{n}})(\mathbf{m})
    = \sum_{j_s\leq k_s} \sum_{\nu_s=0}^\infty
    \frac{g_{j_sk_s}(\nu_s;\mathbf{m})}{2(|{\mathbf n}| - |{\mathbf
    m}|)} \sum_{j_{s-1}\leq k_{s-1} } \sum_{\nu_{s-1} =0}^\infty
    \frac{g_{j_{s-1}k_{s-1}}(\nu_{s-1};\mathbf{m}+
    \mathbf{E}_{j_sk_s}^{\nu_s})}{2(|{\mathbf n}| - |{\mathbf
    m}+\mathbf{E}_{j_sk_s}^{\nu_s}|)} \\ \times \cdots \sum_{j_1\leq
    k_1} \sum_{\nu_1=0}^\infty
    \frac{g_{j_1k_1}(\nu_1;\mathbf{m}+\sum_{\ell=2}^{s}\mathbf{E}_{j_\ell
    k_\ell}^{\nu_1})}{2(|{\mathbf n}| - |{\mathbf m}+\sum_{\ell=2}^{s}
    \mathbf{E}_{j_\ell k_\ell}^{\nu_\ell}|)} \,
    \delta_{\mathbf{n}}(\mathbf{m} + \mbox{$\sum_{r=1}^{s}
    \mathbf{E}_{j_rk_r}^{\nu_r}$} ) \\ =\sum_{j_1\leq k_1}\cdots
    \sum_{j_s\leq k_s}\sum_{\nu_1,\ldots,\nu_s} \prod_{r=1}^s
    \frac{g_{j_r k_r}(\nu_r;\mathbf{n}
    -\sum_{\ell=1}^r\mathbf{E}_{j_rk_r}^{\nu_r})}{2(|{\mathbf n}| -
    |{\mathbf n}-\sum_{\ell=1}^r\mathbf{E}_{j_\ell
    k_\ell}^{\nu_\ell}|)} \, \delta_{\mathbf{n}}(\mathbf{m} +
    \mbox{$\sum_{r=1}^{s}$} \mathbf{E}_{j_rk_r}^{\nu_r} )
\end{split} 
\end{equation*}
for all $\mathbf{m}\Le\mathbf{n}$. We now observe that
\begin{equation*}
2(|{\mathbf n}| - |{\mathbf
  n}-\mbox{$\sum_{\ell=1}^r$}\mathbf{E}_{j_\ell k_\ell}^{\nu_\ell}|) =
  4r
\end{equation*}
and thus obtain (\ref{alphanm}).  

\subsection{Properties of the reduced eigenfunctions}\label{sec34}
There remains to prove that the reduced eigenfunctions
$P_{\mathbf{n}}$ in Theorem~\ref{mainthm} indeed are well-defined
symmetric polynomials. We do this in three steps: we first establish
that the functions $f_{\mathbf{n}}$ are symmetric polynomials, then,
that the $P_{\mathbf{n}}$ are finite linear combinations of the
functions $f_{\mathbf{n}}$, and finally, that all the expansion
coefficients $\alpha_{\mathbf{n}}(\mathbf{m})$ are well-defined.

A proof of the first fact can be found in \cite{Lang}, but for the
convenience of the reader we give the complete argument.

\begin{lemma}\label{symPolLemma}
The functions $f_{\mathbf{n}}$ are homogeneous symmetric polynomials
of degree $|\mathbf{n}|$ and non-zero only if
\begin{equation*}
  n_j + \ldots + n_N\geq 0,\quad\forall j=1,\ldots,N-1.
\end{equation*}
For each $\mathbf{n}\in\mathbb{Z}^N$ and each partition $\mathbf{m}$,
let
\begin{equation*}
    p_{\mathbf{n}\mathbf{m}} = \sum\prod_{i<j
    }\prod_{r,s=1}^N (-1)^{\kappa_{ij}+\nu_{rs}}
    \binom{\lambda}{\kappa_{ij}} \binom{-\lambda}{\nu_{rs}},
\end{equation*}
where the second sum extends over all non-negative integers
$\kappa_{ij}$ and $\nu_{rs}$ such that
\begin{equation*}
  m_j = \sum_{l=1}^N\nu_{jl}\quad\textrm{and}\quad n_j =
  \sum_{l=1}^{j-1}\kappa_{lj} - \sum_{l=j+1}^N\kappa_{jl} +
  \sum_{l=1}^N\nu_{lj}.
\end{equation*}
Then
\begin{equation}
\label{monomialDecomp}
  f_{\mathbf{n}} = \sum_{|\mathbf{m}|=|\mathbf{n}|}
  p_{\mathbf{n}\mathbf{m}} M_{\mathbf{m}}.
\end{equation}
\end{lemma}

\begin{proof}
Since $|y_j|<|y_k|$ and $|x_j|<|y_k|$ along the integration paths in
$f_{\mathbf{n}}$, the terms in the fraction contained in its integral
kernel can be expanded in binomial series in $y_j/y_k$ and $x_j/y_k$,
respectively. The integrals can then be computed using the residue
theorem, and this yields
\begin{equation*}
  \begin{split}
    f_{\mathbf{n}}(\mathbf{x}) = & \sum(-1)^{\kappa_{ij}+\nu_{rs}}
    \binom{\lambda}{\kappa_{ij}}\binom{-\lambda}{\nu_{rs}}x_r^{\nu_{rs}},
  \end{split}
\end{equation*}
where the last sum is to be taken over all non-negative integers
$\kappa_{ij}$ and $\nu_{rs}$ such that
\begin{equation}
\label{constrEq}
  n_j - \sum_{l=1}^{j-1}\kappa_{lj} +
  \sum_{l=j+1}^N\kappa_{jl} - \sum_{l=1}^N\nu_{lj} = 0.
\end{equation}
Recalling the definition of the monomials $M_{\mathbf{m}}$ we deduce
(\ref{monomialDecomp}). To prove that the functions $f_{\mathbf{n}}$
are homogeneous of degree $|\mathbf{n}|$ note that the degree of each
monomial $M_{\mathbf{m}}$ in the decomposition in
(\ref{monomialDecomp}) is given by
\begin{equation*}
  |\mathbf{m}| = \sum_{j,l=1}^N\nu_{jl} = \sum_{j=1}^Nn_j\equiv
  |\mathbf{n}|.
\end{equation*}
It remains only to prove that the functions $f_{\mathbf{n}}$ are
polynomials, i.e., that (\ref{constrEq}) only has a finite number of
solutions. To this end consider the equation for $j=N$,
\begin{equation*}
  n_N = \sum_{l=1}^{N-1}\kappa_{lN} + \sum_{l=1}^N\nu_{lN}.
\end{equation*}
It is clear that it only has a finite number of solutions for each
fixed $n_N$. Now observe that, for $j<N$, the possible values of
$\kappa_{jl}$, $l\geq j+1$, are determined by the equations with
larger values of $j$. Also observe that the equation for each fixed
set of $\kappa_{jl}$, $l\geq j+1$, and $n_j$ has only a finite number
solutions. The statement thus follows by induction in $j$, starting
with $j = N$.
\end{proof}

Observe that either the symmetric polynomial $f_{\mathbf{m}}$ or the
coefficient $\alpha_{\mathbf{n}}(\mathbf{m})$ is zero unless
$m_j+\cdots m_N\geq 0$ for all $j=1,\ldots,N-1$ and
$\mathbf{m}\preceq\mathbf{n}$. Clearly only a finite number of
$\mathbf{m}\in\mathbb{Z}^N$ fulfill these conditions. This proves the
second fact: the reduced eigenfunctions are finite linear combinations
of the symmetric polynomials $f_{\mathbf{n}}$. Also note that all sums
in the explicit representation in (\ref{alphanm}) of the coefficients
$\alpha_{\mathbf{n}}(\mathbf{m})$ truncate after a finite number of
terms, and that they therefore are finite. It follows that the reduced
eigenfunctions indeed are well-defined symmetric polynomials. This
concludes the proof of Theorem \ref{mainthm}.

\subsection{The one-particle case}\label{sec35}
It is interesting to note that Theorem~\ref{mainthm} is non-trivial
already in the simplest case $N=1$.  Since the Calogero model for
$N=1$ reduces to the harmonic oscillator with well-known
eigenfunctions given by the Hermite polynomials $H_n$ (see e.g.\
22.6.20 in \cite{Abr}), Theorem~\ref{mainthm} in this case implies
that the functions $P_n$ in (\ref{Pn}) are equal to the Hermite
polynomials up to normalization. Comparing with the standard
definition of the Hermite polynomials in (\ref{Hermite}) we obtain
that
\begin{equation*}
  H_n(x) = \frac{2^n n!}{(\lambda)_n} \sum_{s=0}^\infty (-1)^s \frac{
  (n+\lambda-2s)_{2s}}{4^s s!}\oint_{|y|>|x|} \frac{dy}{2\pi\ii y}
  y^{n-2s} \frac{1}{(1 - x/y)^{\lambda}},
\end{equation*} 
where $(z)_n$ denotes the Pochhammer symbol
\begin{equation*}
  (z)_0 = 1,\quad (z)_n = z(z + 1)\ldots (z + n - 1),
\end{equation*}
defined for $z\in\mathbb{C}$ and $n\in\mathbb{N}_0$. Note that the
series above truncates after a finite number of terms and thus is
well-defined. This identity has an interesting interpretation. Observe
that the Hermite polynomials can be generalized to arbitrary complex
parameters $\nn$ as follows:
\begin{equation*}
  H_\nn(x) : = \sum_{s=0}^\infty (-1)^s \frac{ (\nn
  -2s+1)_{2s}}{s!}(2x)^{\nn -2s}. 
\end{equation*}
It is straightforward to verify that this series reduces to a Hermite
polynomial when $\nn$ is a non-negative integer, and that it satisfies
the Hermite differential equation $(\partial_x^2 -2x\partial_x +2\nn)
H^{\phantom\prime}_\nn(x)=0$ in the sense of formal Laurent
series. However, it is important to note that the series defining
$H_\nn(x)$ does not converge anywhere in the complex plane but is only
asymptotic unless $\nn$ is a non-negative integer.  Using this formal
Laurent series we can formally rewrite the above identity as follows:
\begin{equation*}
  H_n(x) = \frac{n!}{2^{\lambda-1} (\lambda)_n} \oint_{|y|>|x|}
  \frac{dy}{2\pi\ii}\; H_{n+\lambda-1}(x)\frac{1}{(y - x)^{\lambda}}
\end{equation*} 
for any complex $\lambda$. For non-integer $\lambda$ the r.h.s.\ can 
be made well-defined by exchanging the order of integration and
summation.  For integer $\lambda=m+1>2$ we can use the residue theorem
to compute the integrals and recover the well-known identity
\begin{equation*}
H_n(x) = \frac{n!}{2^m (m+n)!}\frac{d^m}{dx^m}H_{n+m}(x)
\end{equation*} 
obeyed by the Hermite polynomials, and we therefore obtained an
interesting generalization of this to the cases when $n$ is not a
non-negative integer.  The integral transforms in our identity looks
like a fractional integral transform which, as is well-known, shift
parameters of hypergeometric functions; see e.g.\ Chapter~13 in
\cite{E2}. However, the details of our identity seem different.

\subsection{Alternative formulas for the eigenfunctions}\label{sec36} 
The results above can now be used to construct another explicit series
representation for the eigenfunctions of the $N$-body Calogero
model. For that we find it convenient to use a somewhat different
normalization for the formal Laurent series $H_\nn$,
\begin{equation}\label{formalEig}
  p_{\nn}(x) = 2^{-\nn}H_{\nn}(x) = \sum_{\kk=0}^{\infty}
  \cc_{\kk}(\nn) x^{\nn-2\kk},\quad \cc_{\kk}(\nn)= (-1)^\kk
  \frac{(\nn-2\kk+1)_{2\kk}}{4^{\kk}\kk!} .
\end{equation}
The idea is to apply a particular integral transform to the identity
in (\ref{FId}) which differs from the one in Section~\ref{sec32} in
that the simple powers $y_j^{\tilde n_j}$ in (\ref{ntilde}) are
replaced by the formal Laurent series $p_{\tilde n_j}(y_j)$. This
leads to recursion relations which are somewhat different from those
in the proceeding discussion but also can be solved explicitly. The
advantage is that this recursion becomes trivial in the free case
$\lambda=0$ but, as we will see, it becomes somewhat more complicated
to deduce.

To obtain the recursion relations we need an explicit formula for
$x^np_{\nn}(x)$, $n\in\mathbb{Z}$, as a linear combination of
$p_{\nn'}(x)$, $\nn'\leq \nn+n$.

\begin{lemma}\label{multLemma}
Let $\mm\in\mathbb{Z}$ and $\nn\in\mathbb{C}$. Then
\begin{equation}
\label{multSeries}
  x^{\mm} p_{\nn}(x) = \sum_{\kk=0}^{\infty}b_\kk(\mm , \nn)p_{\nn+\mm
  -2\kk}(x)
\end{equation}
with 
\begin{equation*}
  \begin{split}
    b_{0}(\mm ,\nn) &= \cc_0(\nn) = 1,\\ b_\kk(\mm ,\nn) &=
    \cc_{\kk}(\nn) - \sum_{j=0}^{\kk-1}b_j(\mm
    ,\nn)\cc_{\kk-j}(\nn+\mm -2j),\quad \kk>0.
  \end{split}
\end{equation*}
and $\cc_{\kk}(\nn)$ as defined in (\ref{formalEig}).
\end{lemma}

\begin{proof}
Observe that the definition of $p_{\nn}$
implies that
\begin{equation*}
  x^{\mm} p_{\nn}(x) = \sum_{\kk=0}^{\infty}\cc_{\kk}(\nn)x^{\nn+\mm
  -2\kk}
\end{equation*}
in the sense of formal Laurent series.  It follows that
\begin{equation*}
  \begin{split}
    x^{\mm} p_{\nn}(x) &= \cc_0(\nn)p_{\nn+\mm }(x) +
    \sum_{\kk=1}^{\infty}\left( \cc_{\kk}(\nn) -
    \cc_0(\nn)\cc_{\kk}(\nn+\mm )\right) x^{\nn+\mm -2\kk}\\&=
    b_0(\mm ,\nn)p_{\nn+\mm }(x) +
    \sum_{\kk=1}^{\infty}\left( \cc_{\kk}(\nn) -
    b_0(\mm ,\nn)\cc_{\kk}(\nn+\mm )\right)  x^{\nn+\mm -2\kk}.
  \end{split}
\end{equation*}
The statement now follows by repeating this procedure of breaking off
the leading term.
\end{proof}

\begin{remark}
Although it is not evident from the statement and proof of
Lemma~\ref{multLemma}, the series in (\ref{multSeries}) truncates for
non-negative integers $\mm$ at $\kk = \mm$, i.e.,
\begin{equation*}
  b_\kk(\mm,\nn) = 0,\quad \kk > \mm , \quad \mm\in\mathbb{N}_0.
\end{equation*}
This can be deduced by observing that differential
equation solved by $p_{\nn}$ implies the
three term recursion relation
\begin{equation*}
  2 p_{\nn+1}(x) - 2 xp_{\nn}(x) + \nn p_{\nn-1}(x) = 0.
\end{equation*}
From this the truncation of the series in (\ref{multSeries}) for
non-negative integers $\mm$ follows. This also shows that the series
does not truncate for negative integers $\mm$.
\end{remark}

Explicit formulas for the coefficients $b_\kk(\mm,\nn)$ can now be
obtained by solving the recursion relations in Lemma \ref{multLemma}.

\begin{cor}\label{multCor}
Let $\kk\in\mathbb{N}_0$, $\mm\in\mathbb{Z}$ and
$\nn\in\mathbb{C}$. Then
\begin{equation}
\begin{split}
    b_\kk(\mm,\nn) &= (-1)^\kk \frac{(\nn-2\kk+1)_{2\kk}}{4^\kk
    \kk!}\\&\quad + \sum_{j=1}^\kk(-1)^{j+\kk} \sum_{0\leq
    \kk_j<\cdots<\kk}
    \frac{(\nn-2\kk_j+1)_{2\kk_j}(\nn+\mm-2\kk+1)_{2(\kk-\kk_j)}}
    {4^\kk\kk_j!(\kk_{j-1}-\kk_j)!\cdots(\kk-\kk_1)!}.\label{bbb}
\end{split}
\end{equation}
\end{cor}

\begin{proof}
Lemma \ref{multLemma} implies that
\begin{equation*}
  \begin{split}
    b_\kk(\mm,\nn) =&\ \cc_{\kk}(\nn) +
    \sum_{j=1}^{\kk}(-1)^j\sum_{\kk_1=0}^{\kk-1}
    \sum_{\kk_2=0}^{\kk_1-1}\cdots
    \sum_{\kk_j=0}^{\kk_{j-1}-1}\cc_{\kk_j}(\nn)\\&\
    \cc_{\kk_{j-1}-\kk_j}(\nn+\mm-2\kk_j)\ldots
    \cc_{\kk-\kk_1}(\nn+\mm-2\kk_1).
  \end{split}
\end{equation*}
The statement is now obtained by using the explicit form of the
coefficients $\cc_{\kk}(\nn)$ and simple properties of the Pochhammer
symbol.
\end{proof}

Suppose that $y_N>y_{N-1}>\ldots>y_1>\max_k(x_k)$. It is then clear
from the previous discussion that the product
\begin{equation*}
  \prod_{j=1}^Np_{\tilde n_j}(y_j)y_j^{\lambda(j-N-1)}
  \frac{\prod_{j<k}(1-y_k/y_j)^{\lambda}}
  {\prod_{j,k=1}^N(1-x_k/y_j)^{\lambda}}
\end{equation*}
is a formal multi-variable Laurent series in the variables
$y_j$. Appealing to the residue theorem we define
\begin{equation}
\label{altfn}
  f^{\HH}_{\mathbf{n}}(\mathbf{x}):=
  \prod_{j=1}^N\left(\oint_{\mathcal{C}_j}\frac{dy_j}{2\pi\ii y_j}y_j
  p_{\tilde n_j-1}(y_j)y_j^{\lambda(j-N-1)}\right)\frac{\prod_{
  j<k}(1-y_j/y_k)^{\lambda}}
  {\prod_{j,k}(1-x_j/y_k)^{\lambda}}
\end{equation}
as the coefficient of the term $(y_1\ldots y_N)^{-1}$ in this Laurent
series; we use the superscript `$\HH$' to distinguish these from the
analog functions defined in Section~\ref{sec32}. This prescription
amounts to interchanging the integrations and summations.  Following
the proof of Lemma~\ref{symPolLemma} it is readily verified that the
$f^{\HH}_{\mathbf{n}}$ are well-defined symmetric polynomials.
Similarly as in Section~\ref{sec33} we now obtain that
\begin{equation*}
  \begin{split}
    H\hat F^{\HH}_{\mathbf{n}} = & E_{\mathbf{n}}\hat
    F^{\HH}_{\mathbf{n}} +
    2\lambda(\lambda-1)\sum_{j<k}\sum_{\nu=1}^{\infty}
    \nu\sum_{r,s=0}^{\infty} b_r(\nu-1,\tilde n_j) b_s(-\nu-1,\tilde
    n_k)\\&\times\hat
    F^{\HH}_{\mathbf{n}-(1+2r-\nu)\mathbf{e}_j-(1+2s+\nu)\mathbf{e}_k}
  \end{split}
\end{equation*}
for $\hat F^{\HH}_{\mathbf{n}}(\mathbf{x})=\psi_0(\mathbf{x})
f^{\HH}_{\mathbf{n}}(\mathbf{x})$; we used the fact that $hp_{\nu} =
2\nu p_{\nu}$, the functional identity in (\ref{FId}), and
Lemma~\ref{multLemma}. As before we conclude that the action of $H$ on
the functions $\hat F^{\HH}_{\mathbf{n}}$ has triangular structure,
and that there are eigenfunctions of the Calogero model which are of
the form $\psi_{\mathbf{n}}=\beta_{\mathbf{n}}(\mathbf{n})\hat
F^{\HH}_{\mathbf{n}} + \sum_{\mathbf{m}\prec\mathbf{n}}
\beta_{\mathbf{n}}(\mathbf{m})\hat F^{\HH}_{\mathbf{m}}$. The
Schr\"odinger equation $H\psi_{\mathbf{n}} =
E_{\mathbf{n}}\psi_{\mathbf{n}}$ is now implied by the recursion
relations
\begin{equation*}
  \beta_{\mathbf{n}} = \delta_{\mathbf{n}} +
  \SS^{\HH}\beta_{\mathbf{n}},
\end{equation*}
with
\begin{equation*}
  (\SS^{\HH}\beta_{\mathbf{n}})(\mathbf{m}) =
  \frac{1}{E_{\mathbf{n}}-E_{\mathbf{m}}}\sum_{j<k}\sum_{\nu,t,u=0}^{\infty}
  g_{jk}^{tu}(\nu;\mathbf{m})
  \beta_{\mathbf{n}}(\mathbf{m}+\mathbf{E}_{jk}^{tu,\nu}),
\end{equation*}
where we have introduced the notation
\begin{equation}
  g_{jk}^{tu}(\nu;\mathbf{m}) = 2\lambda(\lambda-1)\nu
  b_t(\nu-1;\tilde m_j + 1 + 2t - \nu)b_u(-1-\nu;\tilde m_k + 1 + 2u +
  \nu)\label{ggg}
\end{equation}
and
\begin{equation}
  \mathbf{E}_{jk}^{tu,\nu} = (1+2t-\nu)\mathbf{e}_j +
  (1+2u+\nu)\mathbf{e}_k.\label{EEE}
\end{equation}
Using
\begin{equation*}
  2(|\mathbf{n}| -
  |\mathbf{n}-\mbox{$\sum_{\ell=1}^r\mathbf{E}_{j_\ell
  k_\ell}^{t_\ell u_\ell,\nu_\ell}|$}) = 4\sum_{\ell=1}^r(1+t_l+u_l), 
\end{equation*}
computations similar to the ones in Section~\ref{sec32} lead to the
following result:

\begin{theorem}\label{altFormTheorem}
For $\mathbf{n}\in\mathbb{Z}^N$ let
\begin{equation*}
  P_{\mathbf{n}} = \sum_{\mathbf{m}}
  \beta_{\mathbf{n}}(\mathbf{m})f^{\HH}_{\mathbf{m}}
\end{equation*}
with the functions $f^{\HH}_{\mathbf{m}}$ defined in (\ref{altfn}) and
\begin{equation*}
  \begin{split}
    \beta_{\mathbf{n}}(\mathbf{m}) &= \delta_{\mathbf{n}}(\mathbf{m})
    + \sum_{s=1}^{\infty} \sum_{j_1<k_1}
    \sum_{t_1,u_1,\nu_1=0}^{\infty} \ldots \sum_{j_s<k_s}
    \sum_{t_s,u_s,\nu_s=0}^{\infty}\\&\quad
    \delta_{\mathbf{n}}(\mathbf{m} +
    \mbox{$\sum_{r=1}^s\mathbf{E}_{j_rk_r}^{t_ru_r,\nu_r}$})
    \prod_{r=1}^s \frac{g_{j_rk_r}^{t_ru_r}(\nu_r;\mathbf{n} -
    \sum_{\ell=1}^r\mathbf{E}_{j_\ell k_\ell}^{t_\ell
    u_\ell,\nu_\ell})}{4\sum_{\ell=1}^r(1+t_l+u_l)}
  \end{split}
\end{equation*}
with the quantities given in (\ref{ggg}), (\ref{EEE}) and (\ref{bbb})
above. Then $P_{\mathbf{n}}$ is a reduced polynomial eigenfunction of
the Calogero model corresponding to the eigenvalue $E_{\mathbf{n}}$ in
(\ref{En}).
\end{theorem}

\begin{remark}
The polynomials $P_{\mathbf{n}}$ constructed here should be identical
with the ones obtained in Theorem \ref{mainthm}. We have checked this
for $N=2$ and various cases $\mathbf{n}\in\mathbb{Z}^N$ using
MATHEMATICA.
\end{remark}

\section{Solutions for the remaining classical root systems}\label{sec4}
In this section we show that the constructions of the previous section
can be adapted to the $B_N$ variant of the Calogero model defined by
the Hamiltonian in (\ref{HBN}). The construction is very similar to
the $A_{N-1}$ case, and we therefore are rather sketchy and
concentrate on the necessary changes. To simplify notation we denote
corresponding quantities in the $A_{N-1}$ and $B_N$ cases by the same
symbol, e.g.\ $\psi_0$, $F(\mathbf{x},\mathbf{y})$, $f_{\mathbf{n}}$
etc.\ have a different meaning here and in
Section~\ref{sec3}. However, since the parameter $\mu$ will play a
special role, we will write the $B_N$ Hamiltonian in (\ref{HBN}) as
$H_{\mu}$. Moreover, the analog of the Hermite polynomials are the
Laguerre polynomials denoted by the usual symbol $L^{(a)}_n(x)$; see
e.g.\ \cite{Abr}.

\subsection{A remarkable identity}
The analog of the key identity in Lemma~\ref{FIdLemma} is
\begin{equation}
\label{FIdBN}
  H_{\mu}(\mathbf{x})F(\mathbf{x},\mathbf{y}) = \lbrack
  H_{\lambda-\mu}(\mathbf{y}) + c_N\rbrack F(\mathbf{x},\mathbf{y}),
\end{equation}
where $c_N=2(1 - \lambda)N$ is the same as in the $A_{N-1}$ case but
\begin{equation}
  F(\mathbf{x},\mathbf{y}) = \frac{\prod_{j=1}^N
  e^{-\frac{1}{2}(x_j^2-y_j^2)}x_j^{\mu}
  y_j^{\lambda-\mu}\prod_{j<k}(x_k^2 - x_j^2)^{\lambda}(y_k^2 -
  y_j^2)^{\lambda}}{\prod_{j,k=1}^N(y_k^2 - x_j^2)^{\lambda}}.
\end{equation}
It is important to note that this identity now involves two
Hamiltonians with different coupling parameters $\mu$ and $\lambda-\mu$.

The proof is similar to the one of Lemma~\ref{FIdLemma}: by direct
computation one can check that
\begin{equation}
\label{BNcalH}
\Psi_0(\mathbf{X}) = \prod_{j=1}^\mathcal{N}\left(X_j^{\mu_j}
\ee^{-\frac12 m_j X_j^2} \right)\prod_{j=1}^\mathcal{N}
(X_k^2-X_j^2)^{m_j m_k \mu}
\end{equation}
and the differential operator
\begin{equation*}
  \begin{split}
\mathcal{H} = \sum_{j=1}^\mathcal{N} \left(-\frac1{m_j}\partial_{X_j}^2
+ m_j X_j^2 + \frac{\mu_j(\mu_j-1)}{m_j X_j^2}\right) +
\sum_{j<k}\frac{\lambda}{(X_j^2-X_k^2)^2}
\Bigl( 2(\lambda m_j m_k - 1)\\\times (m_k X_j^2 + m_j
X_k^2)   + (m_k(1+2\mu_j)
-m_j(1+2\mu_k))(X_j^2-X_k^2) \Bigr) 
  \end{split}
\end{equation*}
obey the identity in (\ref{ID}) with 
\begin{equation}
\label{BNcalE0}
\mathcal{E}_0 = \sum_{j=1}^{\mathcal{N}} (1+ 2\mu_j) + 2\lambda \left(
\left(\sum_j m_j\right)^2 - \sum_j m_j^2 \right).
\end{equation} 
The identity to cancel the three-body terms is now
\begin{equation*}
  \frac{X_j^2}{(X_k^2 - X_j^2)(X_j^2-X_l^2)} + \frac{X_k^2}{(X_l^2 -
  X_k^2)(X_k^2-X_j^2)} + \frac{X_l^2}{(X_j^2 -
  X_l^2)(X_l^2-X_k^2)} = 0. 
\end{equation*}
Equation (\ref{FIdBN}) is obtained from this as a special case, as
before.

This also gives, as another important special case, the groundstate
\begin{equation}
    \psi_0(\mathbf{x}) = \prod_{j=1}^N \left( \textrm{e}^{-\frac{1}{2}
    x_j^2}x_j^{\mu} \right) \prod_{j<k}(x_k^2 - x_j^2)^{\lambda}
\end{equation}
and corresponding groundstate energy $E_0$ of the $B_N$ Calogero
model, which are of course well-known; see e.g.\ \cite{OP2}.  We note
in passing that we have obtained a generalization of the $B_N$
Calogero model to particles with different masses, together with its
exact groundstate and groundstate energy.

\subsection{Integral transformation}
The kernel of the integral transform is now taken to be
\begin{equation*}
  \prod_{j=1}^N\frac{\ee^{-\frac{1}{2}y_j^2}y_j^{2\tilde
  n_j+\mu-\lambda}}{2\pi\ii y_j}
\end{equation*}
with the same $\tilde n_j$ as in (\ref{ntilde}). Note that we need to
restrict to even integers $2n_j$ to get non-zero results; see
(\ref{BNfn}) below.

Applying the resulting integral transform to the identity in
(\ref{FIdBN}) straightforward computations lead to
\begin{equation*}
  \begin{split}
    H_{\mu}\hat F_{\mathbf{n}} &= E_{\mathbf{n}}\hat F_{\mathbf{n}} -
    \sum_{j=1}^N 2(2(\tilde n_j+\mu-\lambda)-1)(\tilde n_j-1)\hat
    F_{\mathbf{n}-\mathbf{e}_j}\\&\quad + 4\lambda(\lambda -
    1)\sum_{j<k}\sum_{\nu=1}^{\infty}(2\nu-1)\hat
    F_{\mathbf{n}-(1-\nu)\mathbf{e}_j-\nu\mathbf{e}_k}
  \end{split}
\end{equation*}
with
\begin{equation}
\label{BNEn}
  E_{\mathbf{n}} = 4(n_1 + n_2 +\cdots+ n_N) + E_0,\quad E_0 = N(1 + 2
  \mu + 2\lambda (N-1) )
\end{equation}
and $\hat F_{\mathbf{n}} = \psi_0 f_{\mathbf{n}}$,  
\begin{equation}
\label{BNfn}
  f_{\mathbf{n}}(\mathbf{x}) =
  \prod_{j=1}^N\left(\oint_{\mathcal{C}_j}\frac{dy_j}{2\pi\ii
  y_j}y_j^{2n_j}\right)\frac{\prod_{j<k}(1-y_j^2/y_k^2)^{\lambda}}
  {\prod_{j,k}(1-x_j^2/y_k^2)^{\lambda}}.
\end{equation}
Note that the function $f_{\mathbf{n}}$ by a change of variables,
$x_j^2\to x_j$ and $y_j^2\to y_j$ for all $j=1,\ldots,N$, becomes
identical with the function $f_{\mathbf{n}}$ used in the $A_{N-1}$
case.  Thus Lemma \ref{symPolLemma} directly extends to the $B_N$
case.

\subsection{Construction of eigenfunctions}
The action of the $B_N$ Hamiltonian on the functions $\hat
F_{\mathbf{n}}$ is triangular, which suggests that the Hamiltonian in
(\ref{HBN}) has eigenfunctions of the same form as in (\ref{psin})
with eigenvalues $E_{\mathbf{n}}$ in (\ref{BNEn}). As before, the
Schr\"odinger equation $H\psi_{\mathbf{n}} =
E_{\mathbf{n}}\psi_{\mathbf{n}}$ is implied by the recursion relations
in (\ref{EEal}), where now
\begin{equation}\label{gjkBN}
  \begin{split}
    g_{jk}(\nu;\mathbf{m}) & = 4\lambda(\lambda-1)(2\nu-1)
    (1-\delta_{\nu,0})(1-\delta_{jk})\\&\quad -2(2(\tilde m_j +
    \mu-\lambda)+1)\tilde m_j\delta_{\nu,0}\delta_{jk} 
  \end{split}
\end{equation}
and
\begin{equation}\label{EjkBN}
\mathbf{E}_{jk}^{\nu} = (1-\nu)\mathbf{e}_j +\nu\mathbf{e}_k.
\end{equation} 
With these substitutions the arguments in Section~\ref{sec3} go
through as they stand, and we obtain the analog of Theorem
\ref{mainthm} for the $B_N$ case. To summarize, {\em the reduced
polynomial eigenfunctions of the $B_N$ Calogero Hamiltonian in
(\ref{HBN}) are given by the functions $P_{\mathbf{n}}$ in
Theorem~\ref{mainthm}, where (\ref{fn}), (\ref{gjk}) and (\ref{Ejk})
have to be replaced by (\ref{BNfn}), (\ref{gjkBN}) and (\ref{EjkBN}),
respectively. The corresponding eigenvalues are given by
(\ref{BNEn}).}

\subsection{The one-particle case} The  $B_N$ Calogero Hamiltonian in
(\ref{HBN}) for $N=1$ reduces to $-\partial_x^2 + x^2 +
\mu(\mu-1)x^{-2}$ with well-known eigenfunctions given by the Laguerre
polynomials with non-degenerate eigenvalues; see e.g.\ 22.8.18 in
\cite{Abr}. Comparing this with our eigenfunctions we can conclude
that, in the case $N=1$, the polynomials $P_n(x)$ constructed above
are proportional to the Laguerre polynomials
$L_n^{(\mu-\half)}(x^2)$. By straightforward computations we obtain
that
\begin{equation*}
L^{(\mu-\half)}_n(x) = \ee^{\ii\pi (1-\lambda)}
  \frac{\Gamma(n+\lambda)}{(\lambda)_n} \oint_{|y|>|x|}
  \frac{dy}{2\pi\ii}\; L^{(\mu-\lambda+\half)}_{n+\lambda-1}(x)\frac{1}{(y -
  x)^{\lambda}},
\end{equation*}
where we have made the substitutions $y^2\to y$ and $x^2\to x$ as well
as extended the definition of the Laguerre polynomials to arbitrary
complex values $\nn$ as follows:
\begin{equation*}
L^{(\mu-\half)}_\nn (x) = \sum_{k=0}^\infty  \ee^{\ii\pi(\nn-k) }
\frac{(\nn+\mu+\half-k)_k}{\Gamma(\nn+1-k) k!}\, x^{\nn-k} 
\end{equation*}
with the Gamma function $\Gamma$; note that $L^{(\mu-\half)}_\nn (x)$
is a formal Laurent series obeying the Laguerre differential equation
$(x\partial_x^2 + (\mu+\half-x)\partial_x + \nn
)L_\nn^{(\mu-\half)}(x)=0$, and it reduces to a Laguerre polynomial
when $\nn$ is a non-negative integer; see e.g.\ 22.3.9 in \cite{Abr}.
As in Section~\ref{sec35} the r.h.s.\ in the previous equation has to
be interpreted by exchanging integrations and summations as well as
computing the integrals of the individual terms using the residue
theorem.  For integer values $\lambda=m+1>2$ we recover the well-known
classical identities
\begin{equation*}
L_n^{(\mu-\half)}(x) =
(-1)^m\frac{d^m}{dx^m}L_{n+m}^{(\mu-\half-m)}(x).
\end{equation*}
As in Section~\ref{sec35}, the general case is similar to known
identities involving fractional integral transforms; see e.g.\
Chapter~13 and 16.6.(5) in \cite{E2}.

Using these results it is straightforward to extend our alternative
formulas for the eigenfunctions in Section~\ref{sec36} to the $B_N$
Calogero model.

\section{Concluding remarks}\label{sec5}
In the present paper we extended a solution method for the Sutherland
model \cite{Lang,Lang5} to the $A_{N-1}$ and $B_N$ Calogero models.
Below we discuss various open questions and mention interesting
complimentary results. In particular, in Remark~\ref{rem_overcomplete}
we comment on the issue of whether our solution is complete or not,
and in Remark~\ref{remspc} we point out further identities
generalizing our key results in (\ref{FId}) and (\ref{FIdBN}). In the
concluding Remark~\ref{remInt} we sketch an interesting alternative
interpretation of our method.

\begin{remark}\label{rem_overcomplete}
The main difference between Sutherland's method \cite{Su2} and ours is
that he expands the reduced eigenfunctions of the Calogero model in
the monomials $M_{\mathbf{n}}$ defined in (\ref{Mn}), whereas we
obtain the reduced eigenfunctions as linear combinations of the more
complicated functions $f_{\mathbf{n}}$ in (\ref{fn}).  

The monomials $M_{\mathbf{n}}$ labeled by partitions
$\mathbf{n}=(n_1,n_2,\ldots,n_N)$ are a basis of the space of
symmetric polynomials, and it is therefore obvious that the
eigenfunctions obtained by Sutherland's method are complete.

On the other hand, the functions $f_{\mathbf{n}}$ and also our
eigenfunctions are labeled by unrestricted $N$-tuples $\mathbf{n}\in
\mathbb{Z}^N$, and it therefore seems that we are working with
overcomplete sets of functions. While this is the reason why we can
get more explicit formulas, it also makes the questions of
completeness of our solution more complicated.  It would therefore be
interesting to extend our results to the higher differential operators
commuting with the Hamiltonians since this might shed some light on
this important issue.

It is interesting to note that the very same functions
$f_{\mathbf{n}}$ appeared as building blocks in all the models we so
far have solved by our method: in the $A_{N-1}$ Calogero model they
appear as functions of the variables $x_j$ (see (\ref{fn})), in the
$B_N$ Calogero model as functions of $x_j^2$ (see (\ref{BNfn})), and
for the trigonometric Sutherland case we found the very same functions
but in the variables $z_j=\ee^{\ii x_j}$ \cite{Lang}.

We finally mention that we convinced ourselves that, for $\lambda=1$,
the $f_{\mathbf{n}}$ are up to a difference in sign identical with the
Schur polynomials.
\end{remark}

\begin{remark}\label{remspc} 
The key to our method was the identity in Lemma~\ref{FIdLemma}, but in
its proof we obtained a more general result, given in
(\ref{calH})--(\ref{calE0}), which has other interesting special cases,
as discussed in Remark~\ref{rem2}. We now point out further
interesting identities which can be obtained from this general result.

For example, it is possible to generalize Lemma~\ref{FIdLemma} by
allowing the particle numbers $N$ and $M$ in the $\mathbf{x}$- and
$\mathbf{y}$-variables to be different: choosing ${\mathcal N}=N+M$,
$X_j=x_j$, $m_j=+1$ and $X_{N+k}=y_k$, $m_{N+k}=-1$ for
$j=1,2,\ldots,N$ and $k=1,2,\ldots,M$ we obtain the identity
\begin{equation}
\label{ID1}
H_{N}(\mathbf{x})F_{N,M}(\mathbf{x},\mathbf{y}) = [H_{M}(\mathbf{y}) +
c_{N,M}] F_{N,M}(\mathbf{x},\mathbf{y})
\end{equation}
for the function 
\begin{equation}
\label{FNM}
F_{N,M}(\mathbf{x},\mathbf{y}) = \prod_{j=1}^N \ee^{-\frac12 x_j^2}
\prod_{J=1}^M \ee^{\frac12 y_J^2}\frac{\prod_{1\leq j<k\leq
M}(x_k-x_j)^\lambda\prod_{1\leq J<K\leq
M}(y_K-y_J)^\lambda}{\prod_{j=1}^N\prod_{K=1}^N(y_K-x_j)^\lambda}
\end{equation} 
of $N+M$ variables and the constant 
\begin{equation}
\label{cNM} 
c_{N,M}=\lambda(N-M)^2 + (N+M)(1-\lambda);
\end{equation}
$H_N(\mathbf{x})$ in (\ref{ID1}) is the Calogero Hamiltonian in
(\ref{hamiltonian}) and similarly for $H_M(\mathbf{y})$, where we now
also have to indicate the particle numbers and variables.  Another
interesting family of identities is obtained by choosing ${\mathcal
N}=N+M$, $X_j=x_j$, $m_j=+1$ and $X_{N+k}=y_k$, $m_{N+k}=1/\lambda$
for $j=1,2,\ldots,N$ and $k=1,2,\ldots,M$. This yields the identity
\begin{equation*}
H_{N}(\mathbf{x})\tilde F_{N,M}(\mathbf{x},\mathbf{y}) =
[-\lambda H_{M,1/\lambda}(\mathbf{y}) + \tilde
c_{N,M} ] \tilde F_{N,M}(\mathbf{x},\mathbf{y})
\end{equation*}
for the function 
\begin{equation}
\label{tFNM}
\begin{split}
\tilde F_{N,M}(\mathbf{x},\mathbf{y}) = \prod_{j=1}^N \ee^{-\frac12
x_j^2} \prod_{J=1}^M \ee^{-\frac1{2} y_J^2/\lambda}\prod_{1\leq
j<k\leq M}(x_k-x_j)^\lambda \\ \times \prod_{1\leq J<K\leq
M}(y_K-y_J)^{1/\lambda}\prod_{j=1}^N\prod_{K=1}^N(y_K-x_j)
\end{split} 
\end{equation} 
and the constant 
\begin{equation}
\label{tcNM}
\tilde c_{N,M}=(N\lambda + M/\lambda )^2 + N(1-\lambda)+M(1-
1/\lambda );
\end{equation}
the Calogero Hamiltonian for the variables $\mathbf{y}$ is now given
by
\begin{equation}
\label{thamiltonian}
H_{M,1/\lambda}(\mathbf{y}) = \sum_{J=1}^N\left(-\partial_{y_J}^2 +
   y_J^2/\lambda^2 \right) + 2 (1/\lambda)((1/\lambda)
  -1)\sum_{J<K}\frac{1}{(y_J - y_K)^2}. 
\end{equation}
We thus recover a well-known duality between Calogero models with
reciprocal coupling constants.

It is interesting to note that corresponding identities also exist in
the $B_N$ case, but we do not write them down since they can be
obtained from our general result quoted in
(\ref{BNcalH})--(\ref{BNcalE0}) in the same way as explained in the
$A_{N-1}$ case: to obtain the analog of (\ref{FNM})--(\ref{cNM}) fix
the parameters as in the $A_{N-1}$ case and, in addition, set
$\mu_j=\mu$, $\mu_{N+k}=\lambda-\mu$, and for the analog of
(\ref{tFNM})--(\ref{thamiltonian}) the additional parameters are to be
fixed as $\mu_j=\mu$, $\mu_{N+k}=\frac12(3/\lambda-1)$.

With these identities one can obtain many more explicit formulas for
the eigenfunctions of the Calogero models using our method. It is
interesting to note that similar identities were previously found also
in the Sutherland model and its elliptic generalization using quantum
field theory techniques \cite{Lang4}, and we believe that such
identities should exist also for other integrable many-body systems.
\end{remark}

\begin{remark}\label{remInt}
We now sketch an interesting alternative interpretation of our method
which, when explored in more detail, could shed more light on the
questions discussed in Remark~\ref{rem_overcomplete}.\footnote{We
would like to thank Vadim Kuznetsov for useful discussions on this
point.} Suppose that we want to construct eigenfunctions of the
Calogero Hamiltonian in (\ref{hamiltonian}) of the following form
\begin{equation*}
\chi_{\mathbf{n}} = \chi_0 \PQ_{\mathbf{n}},\quad \chi_{0}(\mathbf{x})
= \prod_{j=1}^N \ee^{-\frac12 x_j^2}
\end{equation*}
with $\PQ_{\mathbf{n}}$ linear
combinations of arbitrary monomials
\begin{equation*}
\mathbf{x}^{\tilde{\mathbf{n}}} : = x_1^{\tilde n_1}x_2^{\tilde
n_2}\cdots x_N^{\tilde n_N},
\end{equation*}
where the $\tilde n_j = n_j + s_j$ are integers shifted by common
amounts $s_j$ which, at this point, are arbitrary.  If we ignore the
issue of square integrability it is easy to construct such
eigenfunctions: with the ansatz
\begin{equation*}
\PQ_{\mathbf{n}}(\mathbf{x}) = \mathbf{x}^{\tilde{\mathbf{n}}} +
\sum_{\mathbf{m}\Le\mathbf{n}} \alpha_{\mathbf{n}}({\mathbf{m}})
\mathbf{x}^{\tilde{\mathbf{m}}}
\end{equation*}
we find by straightforward computations that the Schr\"odinger
equation $H\chi_{\mathbf{n}} = \check E_{\mathbf{n}}\chi_{\mathbf{n}}$
is equivalent to $\check E_{\mathbf n} = |\mathbf{n}| + |\mathbf{s}|$,
$|\mathbf{s}|=s_1+s_2+\ldots+s_N$, and the recursion relations in
(\ref{EEal}) for the coefficients
$\alpha_{\mathbf{n}}({\mathbf{m}})$. As shown in Section~\ref{sec33},
these relations can be easily solved, and their explicit solution is
given in Theorem~\ref{mainthm}. It is important to note that the
construction of these eigenfunctions $\chi_{\mathbf{n}}$ is not
restricted to the Calogero model, but one can easily generalize it to
construct similar eigenfunctions for non-integrable systems like the
generalized Calogero model where the particles have different
masses. However, these eigenfunctions are completely uninteresting
from a physical point of view: the series in the previous formula do
not converge and are only asymptotic. The fact that makes the Calogero
model special is that {\em there exists an operator $\mathcal{F}$
which maps these unphysical eigenfunctions to physical ones}, and this
operator is given by the function $F(\mathbf{x},\mathbf{y})$ in
Lemma~\ref{FIdLemma} in the following sense: for an asymptotic series
\begin{equation*}
\chi(\mathbf{x}) = \chi_0(\mathbf{x})
\sum_{\mathbf{m}\Leq\mathbf{n}} \alpha(\mathbf{m})
\mathbf{x}^{\tilde{\mathbf{m}}}
\end{equation*}
let
\begin{equation*}
\mathcal{F}(\chi)(\mathbf{x}) = \sum_{\mathbf{m}} \alpha(\mathbf{m})
\prod_{j=1}^N \left( \oint_{\mathcal{C}_j} \frac{dy_j}{2\pi\ii y_j}
\right)F(\mathbf{x},\mathbf{y}) \chi_0 (\mathbf{y}) 
\mathbf{y}^{\tilde{\mathbf{m}}}
\end{equation*}
with the integration paths defined in (\ref{Cj}). Note that this map
is well-defined if we set $s_j= \lambda(N+1-j)$, and then the
functions
\begin{equation*}
\mathcal{F}(P_\mathbf{n}\chi_0) = \psi_0 f_\mathbf{n}
\end{equation*}
are equal to the building blocks of our solution given in 
(\ref{groundState}) and (\ref{fn}).
Moreover, since Lemma~\ref{FIdLemma} implies that
\begin{equation*}
H  \mathcal{F}(\chi)= \mathcal{F}([H+c_N]\chi), 
\end{equation*}
we conclude that 
\begin{equation*}
\psi_\mathbf{n} : = \mathcal{F}(\chi_\mathbf{n})
\end{equation*}
is an eigenfunction of the Calogero Hamiltonian with eigenvalue
$E_{\mathbf n}=\check E_{\mathbf n}+c_N$.

Obviously the generalized identities pointed out in
Remark~\ref{remspc} provide operators which, similarly, transform
unphysical eigenfunctions of the $M$-variable Calogero model to
physical $N$-variable ones, and thus our results in this paper give
the further explicit formulas for the eigenfunctions mentioned in
Remark~\ref{remspc}.

We note that the operator $\mathcal{F}$ is similar to the $Q$-operator
which has appeared in a separation-of-variables approach to the
Sutherland model \cite{KMS}.

\end{remark}

\subsection*{Acknowledgments} 
We are grateful F.\ Calogero, P.\ Forrester and V.\ Kuznetsov for
helpful discussions.  This paper was written in part during a
scientific gathering at the {\it ``Centro Internacional de Ciencias
A.C.''} (CIC) in Cuernavaca (Mexico) devoted to Integrable Systems,
and we would like to thank Francesco Calogero and Antonio Degasperis
for inviting us there.  This work was supported by the Swedish Science
Research Council (VR), the G\"oran Gustafsson Foundation, and the
European grant ``ENIGMA''.
\begin{appendix}

\section{The relation to Calogero's original model}\label{appA} 
We give here a brief account of the relation between the Hamiltonian in
(\ref{hamiltonian}) and the following one studied by Calogero
\cite{Cal,Cal2}:
\begin{equation*}
  H_{Cal} = -\sum_{j=1}^N\partial_{x_j}^2 + \sum_{j<k}\left( \omega^2
  (x_j - x_k)^2 + 2\lambda(\lambda-1)\frac{1}{(x_j - x_k)^2}\right).
\end{equation*}
It seems well known that they differ only in their center of mass
motion\footnote{We thank F.\ Calogero for explaining this to us.} but
we have been unable to find a discussion of this in the literature.

The simplest way to see this is to verify the following identity:
\begin{equation*}
H_{Cal} + \omega^2(x_1+x_2+\cdots+x_N)^2 = H\; \mbox{ for
$N\omega^2=1$} .
\end{equation*}
Since $H_{Cal}$ is translational invariant it is possible to use
center of mass coordinates with $r_0=(x_1+x_2+\cdots + x_N)/\sqrt N$
and other coordinates $r_j$, $j=1,2,\ldots,N-1$, linearly independent
of $r_0$ and
\begin{equation*}
H_{Cal} = -\frac1N \partial_{r_0}^2 + H_{CM}
\end{equation*}
were $H_{CM}$ only depends on the $r_j$ with $j>0$. We thus conclude that
\begin{equation*}
H = -\frac1N \partial_{r_0}^2 + r_0^2 + H_{CM},
\end{equation*}
where we used that $N\omega^2 = 1$. This makes manifest that the
difference between the two Hamiltonians $H$ and $H_{Cal}$ lies only in
the center of mass motion: in the former it is trapped in a harmonic
oscillator potential, and in the later it is free. Thus the
Hamiltonian $H_{Cal}$ has discrete spectrum only if its center of mass
motion is fixed, i.e., if only $H_{CM}$ is considered. One the other
hand, $H$ has purely discrete spectrum even if the center of mass is
not fixed, and it is therefore simpler to work with.

It is instructive compute $H_{CM}$ explicitly. For $N=3$ the center of
mass coordinates are $r_0$ above and
\begin{equation*}
  r_1 = \frac{1}{\sqrt 2}(x_1 - x_2),\quad r_2 = \frac{1}{\sqrt 6}(x_1
  + x_2 - 2x_3).
\end{equation*}
By a straightforward computation follows that
\begin{equation*}
  H_{CM} = - \partial_{r_1}^2 - \partial_{r_2}^2 + 3 \omega^2(r_1^2 +
  r_2^2) + 2\lambda(\lambda-1)\left( \frac1{2r_1^2} +
  4\frac{r_1^2+3r_2^2}{(r_1^2-3r_2^2)^2} \right) .
\end{equation*}
This computation extends straightforwardly to the general case by a
generalization of the center of mass coordinates to arbitrary $N$. The
latter can be found in Reference~\cite{Cal2}, for example.

\section{Proof of the identity in (\ref{hermId})}\label{appB} 
In this appendix we sketch a proof of the identity in (\ref{hermId}). It
should be noted though that the proof does not give the explicit form
of the coefficients $c_{r,s}$, which seemingly is rather complicated.

The key ingredient in the proof is the following:
 
\begin{lemma}\label{idLemma}
Let $n,m\in\mathbb{N}_0$ and $n\geq m$. Then
\begin{equation*}
  \begin{split}
    \frac{1}{x-y}(\partial_x - \partial_y)(x^ny^m + y^nx^m) = (n -
    m)\sum_{k=1}^{n-m-1}x^{n-1-k}y^{m-1+k}\\ - mx^{n-1}y^{m-1} -
    my^{n-1}x^{m-1}.
  \end{split}
\end{equation*}
\end{lemma}

\begin{proof}
The case $n = m$ is easily verified. Now suppose that $n \geq m +
1$. Then
\begin{equation*}
  \begin{split}
    \frac{1}{x-y}(\partial_x - \partial_y)(x^ny^m + y^nx^m) =
    n\frac{x^my^m}{x-y}(x^{n-m-1} - y^{n-m-1})\\ -
    m\frac{x^{m-1}y^{m-1}}{x-y}(x^{n-m+1} - y^{n-m+1}).
  \end{split}
\end{equation*}
Expand the fractions in geometric series to obtain
\begin{equation*}
  n\sum_{k=0}^{n-m-2}x^{n-2-k}y^{m+k} -
  m\sum_{k=0}^{n-m}x^{n-1-k}y^{m-1+k}.
\end{equation*}
Collect terms of equal degree to deduce the statement.
\end{proof}

The next step is to use the series representation of the Hermite
polynomials and apply Lemma \ref{idLemma} to each term
separately. Recall that, for each $n\in\mathbb{N}_0$, the Hermite
polynomial $H_n$ is a polynomial of degree $n$. Hence, this procedure
clearly gives a polynomial of degree at most $n-1$ in the variables
$x$ and $y$. The validity of the identity in (\ref{hermId}) now
follows from the fact that as $n$ runs through the integers less than
or equal to $m$, for some $m\in\mathbb{N}_0$, the $H_n$ form a basis
for the space of polynomials with degree at most $m$.

\end{appendix}

\end{document}